\documentclass[conference]{IEEEtran}
\IEEEoverridecommandlockouts
\renewcommand{\thesection}{\Roman{section}}
\usepackage{booktabs}
\usepackage{fancyhdr}
\usepackage{algorithm}
\usepackage[usenames, dvipsnames]{xcolor}
\usepackage{algorithmic}
\usepackage{colortbl}
\usepackage{caption}
\usepackage{graphicx}
\usepackage{array}
\usepackage{adjustbox}
\usepackage{hyperref,graphicx,color,float}
\definecolor{myblue}{RGB}{10, 150, 200}

\usepackage{cite}
\usepackage{multirow}
\usepackage{amsmath,amssymb,amsfonts}
\usepackage{algorithmic}
\usepackage{graphicx}
\usepackage{textcomp}
\usepackage{comment}
\definecolor{highlightColor}{HTML}{E6FFE6}

\usepackage{xcolor}
\def\BibTeX{{\rm B\kern-.05em{\sc i\kern-.025em b}\kern-.08em
    T\kern-.1667em\lower.7ex\hbox{E}\kern-.125emX}}

\title{From Code to Career: Assessing Competitive Programmers for Industry Placement}

\author{
\IEEEauthorblockN{ Md Imranur Rahman Akib\textsuperscript{1}, Fathima Binthe Muhammed\textsuperscript{1}, Umit Saha\textsuperscript{1}, Md Fazlul Karim Patwary\textsuperscript{1} \\Mehrin Anannya\textsuperscript{1}, Md Alomgeer Hussein\textsuperscript{2}, Md Biplob Hosen\textsuperscript{*,1,2}}
\IEEEauthorblockA{\textsuperscript{1}\textit{Institute of Information Technology}, \textit{Jahangirnagar University}, Savar, Dhaka, Bangladesh \\
\IEEEauthorblockA{\textsuperscript{2}\textit{University of Maryland Baltimore County}, Maryland, United States }
*Corresponding author [biplob.hosen@juniv.edu]}
}

\begin{document}
\maketitle

\begin{abstract}
In today's fast-paced tech industry, there is a growing need for tools that evaluate a programmer's job readiness based on their coding performance. This study focuses on predicting the potential of Codeforces users to secure various levels of software engineering jobs. The primary objective is to analyze how a user's competitive programming activity correlates with their chances of obtaining positions, ranging from entry-level roles to jobs at major tech companies. We collect user data using the Codeforces API, process key performance metrics, and build a prediction model using a Random Forest classifier. The model categorizes users into four levels of employability, ranging from those needing further development to those ready for top-tier tech jobs. The system is implemented using Flask and deployed on Render for real-time predictions. Our evaluation demonstrates that the approach effectively distinguishes between different skill levels based on coding proficiency and participation. This work lays a foundation for the use of machine learning in career assessment and could be extended to predict job readiness in broader technical fields.
\end{abstract}


\section{Introduction}

The demand for skilled software engineers has increased significantly due to rapid advancements in fields such as artificial intelligence, data science, and software development. In response, companies are increasingly prioritizing practical coding abilities and problem-solving skills over traditional academic credentials \cite{b11}. Competitive programming platforms such as Codeforces, LeetCode, and CodeChef have become key tools for identifying and evaluating technical talent \cite{b2, b26}. Among these, Codeforces stands out for its active global community and wide range of algorithmic challenges.

Participation in these platforms fosters essential skills such as algorithmic thinking, debugging, and coding efficiency—traits highly valued by top tech employers like Google, Microsoft, and Meta \cite{b4}. Research indicates that performance metrics such as contest ratings, participation frequency, and rating progression strongly correlate with better job outcomes \cite{b9, b5}. Universities that incorporate competitive programming into their curricula have also reported higher graduate employment rates \cite{b8}.

Recent works have begun exploring the use of machine learning to predict career trajectories based on competitive programming data \cite{b13}. Models such as Random Forest, Support Vector Machines, and deep learning approaches have demonstrated promising results by analyzing features like rating trends, contest participation, and problem-solving patterns \cite{b15, b16, b14, b7, b1}. These efforts highlight the potential of automated, AI-driven systems to deliver accurate and scalable career predictions.

Despite these advances, challenges remain. Many models struggle with issues such as limited real-world job data, a lack of personalized insights, and class imbalances that favor top performers. Additionally, few systems have been deployed in practical settings, limiting their real-world impact.

This research aims to address these gaps by developing a personalized, data-driven prediction system using Codeforces data. The goal is to identify key indicators of career success and build an interpretable machine learning model that not only forecasts employability but also provides actionable feedback. By bridging the gap between programming performance and career outcomes, this work seeks to support both aspiring programmers and recruiters through a novel approach to talent evaluation.

\section{Methodology}
The proposed methodology consists of several stages: data collection from the Codeforces platform, data preprocessing and dataset splitting for training classification models, model evaluation and performance comparison, and the selection of the best-performing model for deployment. Figure 1 illustrates the overall workflow.
\begin{figure}[htbp]
    \centering
    \includegraphics[width=0.5\textwidth]{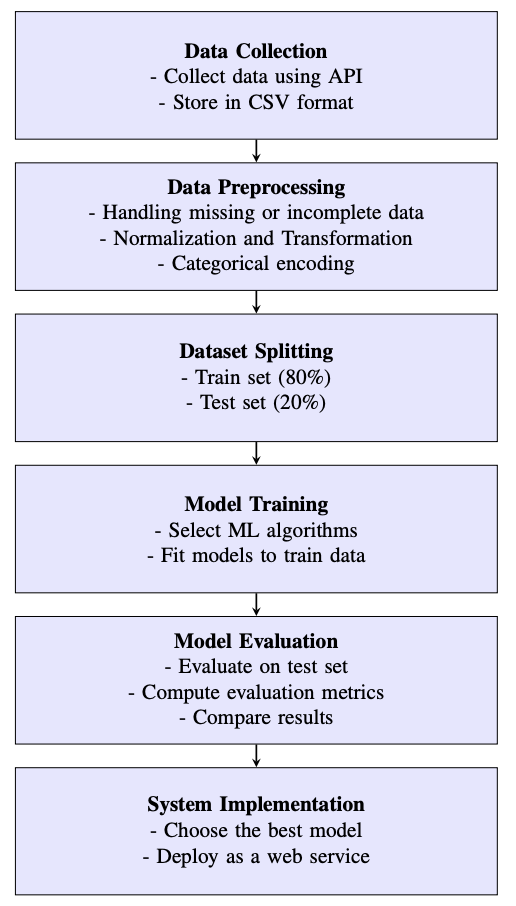} 
    \caption{Workflow diagram of the system.}
    \label{fig:figure1}
\end{figure}

\subsection{Data Collection}
This study collected data from the Codeforces API, which provides structured JSON data on users, contests, problems, and submission history. Key endpoints such as \texttt{user.rating}, \texttt{user.status}, \texttt{contest.list}, and \texttt{problemset.problems} were used to gather information on user ratings, submission outcomes, problem difficulty, and contest metadata.

Codeforces handles were linked with publicly available job status labels to enable supervised learning. For each user, rating histories and submission records were retrieved to compute features related to performance, problem-solving behavior, and engagement. The data was processed using Python to extract relevant metrics, handle missing or invalid entries, and ensure data integrity. API rate limits and occasional downtimes were addressed using delay mechanisms and error-handling strategies.

 The extracted features were grouped into four categories: \textit{performance metrics}, \textit{engagement}, \textit{problem-solving skills}, and \textit{rating trends}. Performance metrics included best rating \cite{b13}, total contests \cite{b20}, total problems solved \cite{b21}, average problem rating \cite{b22}, acceptance ratio \cite{b23}, and ranking statistics \cite{b24}. Engagement was measured through contests per month \cite{b20}, submissions per day \cite{b25}, and days active \cite{b25}. Problem-solving skills were represented by the number of problems solved across difficulty levels and algorithmic categories \cite{b20}. Rating trends, including progression and improvement rate \cite{b13}, captured user growth over time.

This comprehensive feature set enables predictive modeling of career outcomes by capturing users’ technical ability, consistency, and progression in competitive programming. An overview of the dataset is presented in Table 1.

\begin{table}[h]
\centering
\begin{tabular}{lcc}
\toprule
\textbf{Current Job Status} & \textbf{Count} & \textbf{Percentage} \\
\midrule
0 - Needs further practice     & 95  & 15.2\% \\
1 - Entry-level positions       & 331 & 53.0\% \\
2 - Mid-level positions         & 141 & 22.6\% \\
3 - Ready for top tech companies & 58  & 9.3\% \\
\midrule
\textbf{Total}                  & 625 & 100\% \\
\bottomrule
\end{tabular}
\caption{Programmers job status distribution.}
\label{tab:job_status_distribution}
\end{table}

\subsection{Data Prepocessing}
Effective data processing is crucial for building robust machine-learning models. This study addressed two significant aspects: handling missing data and normalizing features. The dataset, sourced from the Codeforces API, included gaps due to user inactivity, incomplete submissions, and API failures. To ensure integrity, strategies such as default value assignment (e.g., zero or NaN), median imputation for problem ratings, and linear interpolation for rating history were employed. Records lacking meaningful activity were excluded to reduce noise and bias.

To prepare the data for model training, normalization and transformation techniques were applied. Features with large numerical ranges (e.g., best rating) were scaled using Min-Max normalization, while others with high variability (e.g., average rank) were standardized using Z-score normalization. Log transformation was used to reduce skewness in features like 'problems solved' and 'submissions per day.' Categorical variables, such as 'problem tags' and 'job status,' were encoded using One-Hot Encoding and Label Encoding. These preprocessing steps ensured balanced feature contributions, improved model accuracy, and enhanced generalization.

\subsection{Exploratory Data Analysis}
Exploratory Data Analysis (EDA) was conducted to identify patterns and relationships within the dataset before building predictive models. The analysis focused on the performance of Codeforces users, including contest participation, problem-solving skills, and rating trends. Visualizations, such as histograms, scatter plots, bar charts, and correlation heatmaps, revealed that consistent involvement leads to better ratings and that specific problem categories, like dynamic programming and graphs, are more frequently solved. These insights guided feature selection and informed the development of machine-learning models for predicting job readiness.

\begin{figure}[htbp]
    \centering
    \includegraphics[width=0.9\columnwidth]{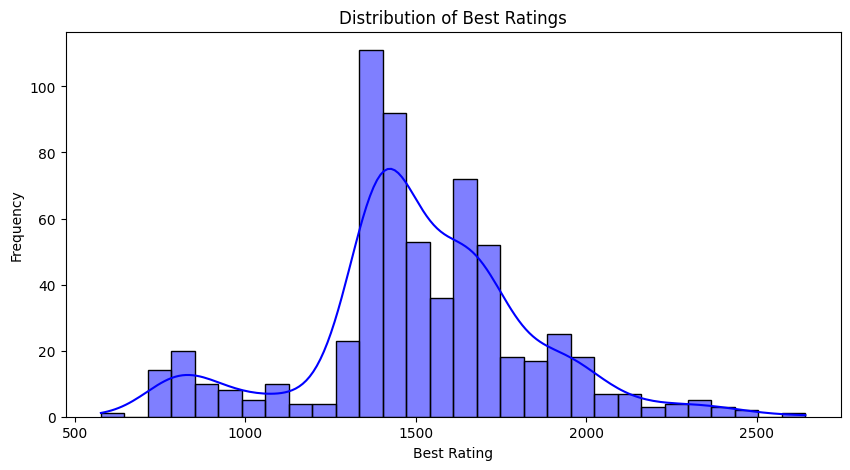}
    \caption{Distribution of best ratings.}
    \label{fig:eda1}
\end{figure}

A histogram [Figure \ref{fig:eda1}] was used to illustrate the distribution of best ratings among Codeforces users. The plot revealed that most users fall within a specific range, with a noticeable skew toward lower ratings. A density estimation line highlighted the most common rating intervals, providing insights into the performance distribution of competitive programmers.

\begin{figure}[htbp]
    \centering
    \includegraphics[width=1\columnwidth]{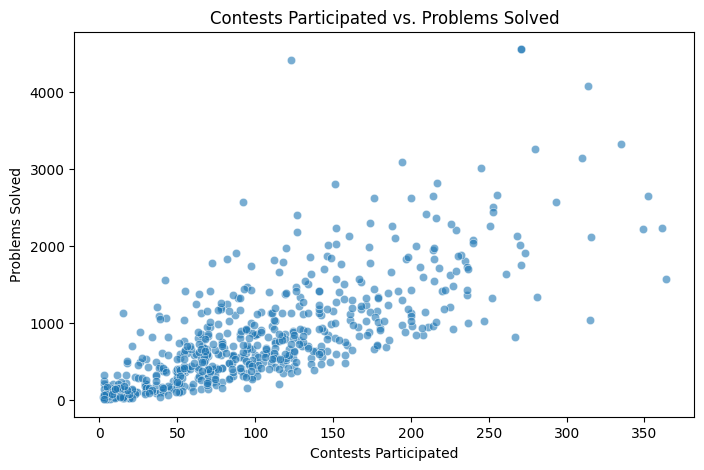}
    \caption{Contests participated vs. problems solved.}
    \label{fig:eda2}
\end{figure}

Figure \ref{fig:eda2} was employed to visualize the correlation between the number of contests participated in and the number of problems solved. The plot showed a clear trend where higher contest participation generally leads to more problems solved, reinforcing the importance of consistent practice in improving competitive programming performance.

\begin{figure}[htbp]
    \centering
    \includegraphics[width=1\columnwidth]{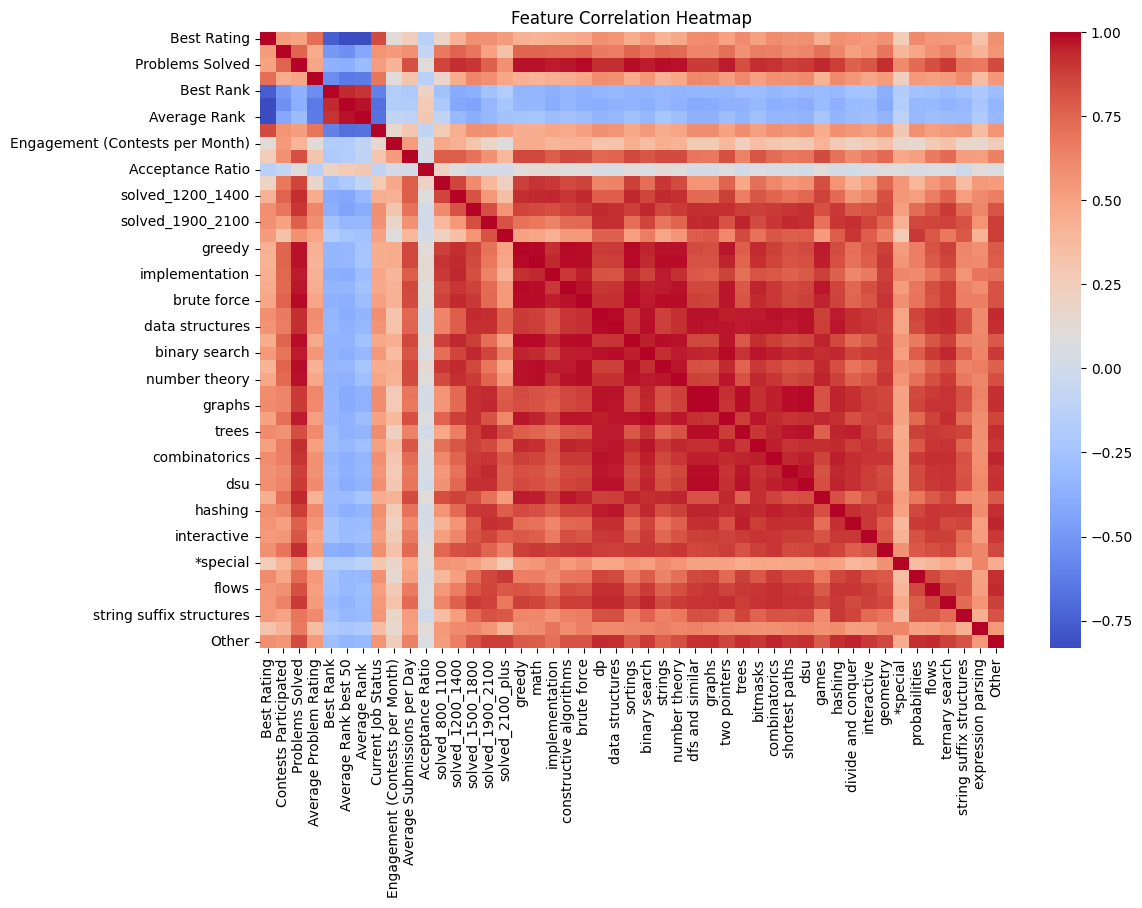}
    \caption{Feature correlation analysis.}
    \label{fig:eda3}
\end{figure}

A hitmap [Figure \ref{fig:eda3}] was used to showcase correlations between numerical features in the dataset. Strong correlations, such as between 'best rating' and 'average problem rating,' indicated dependencies between variables. Understanding these relationships helped refine the machine learning models by identifying the most influential features.

\begin{figure}[htbp]
    \centering
    \includegraphics[width=0.9\columnwidth]{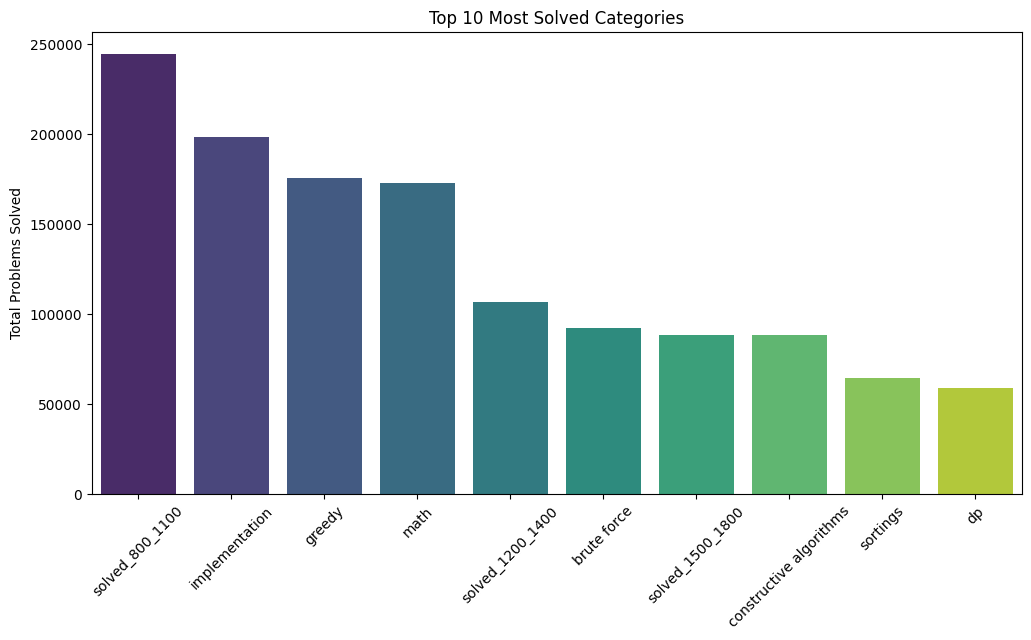}
    \caption{Top 10 problem categories solved.}
    \label{fig:eda4}
\end{figure}

Figure \ref{fig:eda4} highlighted the most popular problem categories among competitive programmers. Categories like 'Greedy,' 'Dynamic Programming,' and 'Implementation' emerged as the most frequently solved, underscoring their relevance in competitive programming and technical interviews. These categories often serve as foundational topics that test a programmer's algorithmic thinking and problem-solving strategies. The high frequency of these problems also suggests their accessibility and widespread inclusion in online judge platforms and contests. Moreover, the diversity of topics within the top 10 indicates a well-rounded approach among programmers aiming to build comprehensive skill sets.

\begin{figure}[htbp]
    \centering
    \includegraphics[width=0.9\columnwidth]{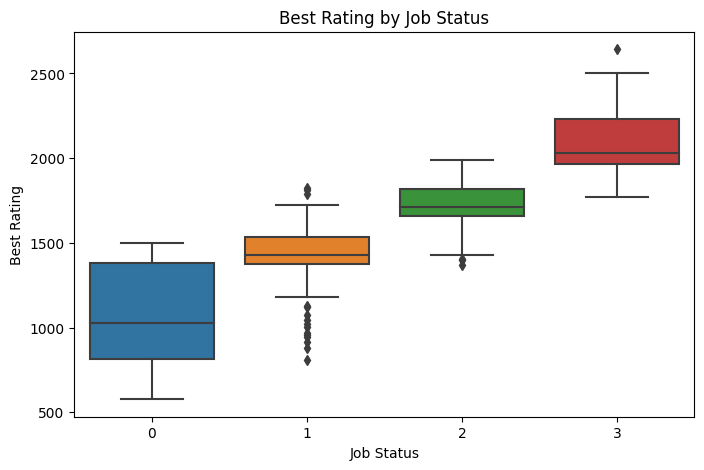}
    \caption{Job status vs. best rating.}
    \label{fig:eda5}
\end{figure}

A box plot [Figure \ref{fig:eda5}] analyzed the variation in 'best rating' across different job statuses. The plot revealed that higher-rated programmers tend to have better job prospects, emphasizing the significance of competitive programming as a measure of job readiness in the tech industry. Median ratings were noticeably higher among employed individuals, suggesting that consistent performance in programming contests can be a strong indicator of practical problem-solving skills. Furthermore, the presence of outliers highlights exceptional performers across all job statuses, indicating that while rating is a strong factor, other elements, such as networking, communication, and domain knowledge, also play a role in employment outcomes.

\begin{figure}[htbp]
    \centering
    \includegraphics[width=.9\columnwidth]{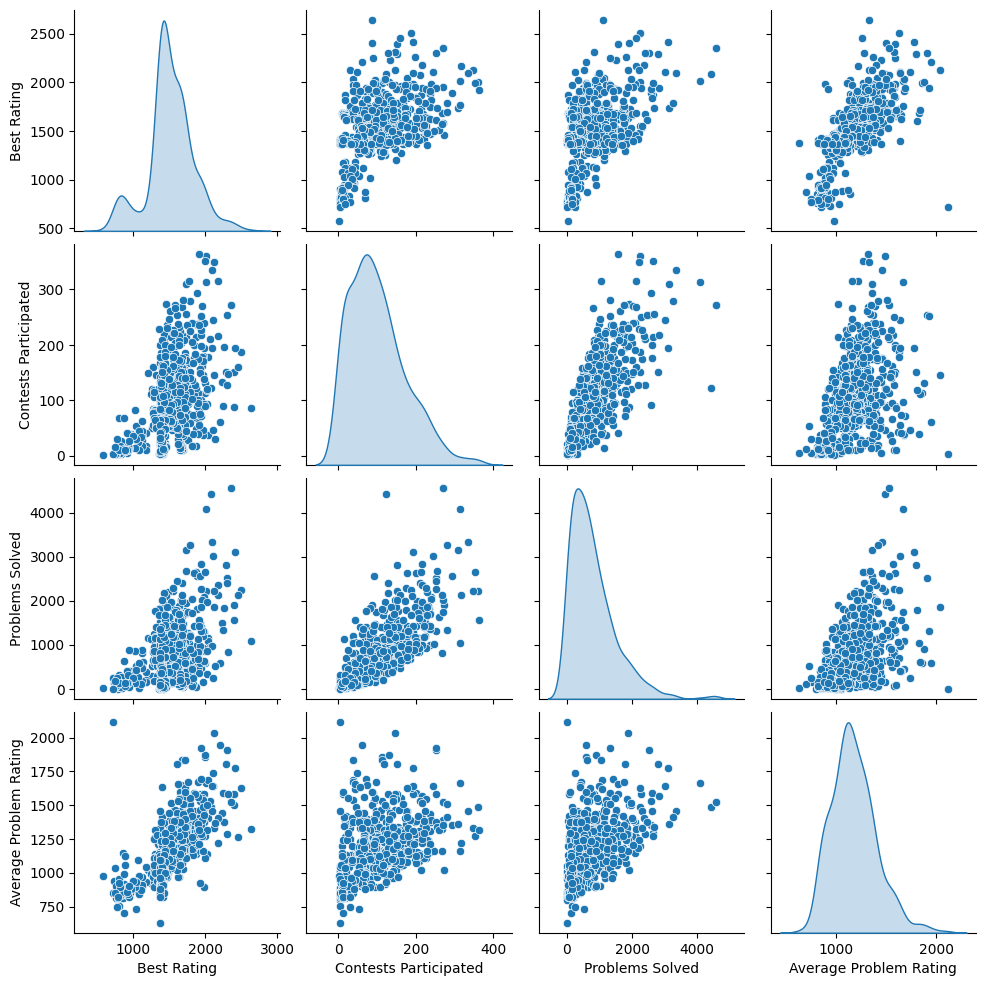}
    \caption{Features relationships.}
    \label{fig:eda6}
\end{figure}

Figure \ref{fig:eda6} provided a visual representation of relationships between key performance metrics. Density plots on the diagonal illustrated the distribution of these attributes, aiding in the selection of features for predictive modeling. This comprehensive visualization helped identify patterns and dependencies, ensuring a robust feature set for the machine learning models. Strong linear and non-linear correlations between features became apparent, revealing potential multicollinearity and highlighting which variables might be most informative for the model. Additionally, the plot made it easier to identify outliers and clustering behavior, both of which are crucial for understanding the data structure before applying any algorithms.

By leveraging these EDA techniques, the study gained valuable insights into the dataset, enabling the development of accurate and reliable predictive models for assessing job readiness based on competitive programming performance.

\subsection{Model Development and Evaluation}
To predict users' job status based on their competitive programming activities, three machine learning models - Random Forest Classifier, Support Vector Machine (SVM), and K-Nearest Neighbors (KNN) are evaluated. 

The Random Forest classifier, an ensemble method, reduces overfitting by aggregating the outputs of multiple decision trees and is particularly effective at handling high-dimensional data with minimal tuning. SVM, known for their robustness in both linear and non-linear classification tasks, perform well in high-dimensional feature spaces by finding an optimal hyperplane that maximizes the margin between classes. KNN, a simple yet intuitive instance-based algorithm, classifies instances based on the majority label of the nearest neighbors but can become computationally expensive as the dataset grows, particularly during inference.

To ensure a comprehensive assessment of model performance, multiple evaluation metrics were used. Precision measures the proportion of accurate optimistic predictions among all predicted positives, indicating how well the model avoids false positives. Recall (or sensitivity) reflects the ability of the model to identify all relevant instances, i.e., the proportion of true positives captured out of all actual positives. The F1-score, the harmonic mean of precision and recall, provides a balance between these two metrics and is especially useful in cases of class imbalance. Accuracy, a more general metric, calculates the proportion of correctly predicted instances out of the total samples. While accuracy provides an overall sense of model correctness, the combination of precision, recall, and F1 score enables a more nuanced evaluation, particularly in imbalanced classification settings. These metrics collectively enabled a thorough comparison of classifier performance in predicting job placement from competitive programming data.

\subsection{System Implementation and Deployment}
The system is implemented using modern software tools and deployed through a cloud-based architecture. Python powers the backend, with HTML, CSS, and JavaScript handling the front end. Scikit-learn is used for training machine learning models, including Random Forest, SVM, and KNN. Flask, a lightweight Python web framework, serves as the backend API, while Material UI and Google Fonts enhance the frontend design. GitHub enables version control and collaboration, and Render is used for continuous deployment.

To ensure reliability, the backend includes robust error handling for API failures, invalid usernames, and unexpected data formats. To handle Codeforces API rate limits (1 request per second), a time.sleep(1) delay is enforced between API calls, preventing service disruptions. The system also validates API responses before proceeding with feature extraction.

For model versioning, the deployed API loads models from a versioned directory structure (e.g., models/v1/, models/v2/), allowing seamless updates and rollback if needed. A symbolic link is used to point to the active version, ensuring zero downtime upgrades. Model metadata (training date, features used, accuracy) is logged for each release.

The front end is responsive and accessible, allowing users to input their Codeforces handle and view color-coded career predictions. The API integrates with the model to return real-time responses. GitHub and Render integration supports automatic redeployment upon code updates, ensuring the live system remains current and stable.

\section{Results and Analysis}
This section presents a comprehensive analysis of the model's performance, encompassing its evaluation using multiple metrics, a comparison with traditional career prediction methods, and insights into its practical applicability in real-world settings.

\subsection{Model Performance on Test Data}
The effectiveness of three machine learning models was evaluated using a the test dataset to ensure unbiased assessment. The models classified users into four career status categories: 0: Needs further practice, 1: On the right track for a job in BD (Entry-level positions), 2: Can get a high-paying job in BD (Mid-level positions), and 3: Ready for top tech companies. Table~\ref{tab:comparison} illustrates a comparison of models outcomes. The Random Forest classifier outperformed SVM and KNN across all metrics, achieving an accuracy of 88.8\%, with balanced precision, recall, and F1 Scores of 0.85. SVM followed with an accuracy of 84.3\%, performing slightly better in recall (0.86) but somewhat lower in precision. KNN underperformed, particularly in recall (0.52), indicating difficulty in correctly classifying minority classes.

\begin{table}[h]
\centering
\small 
\begin{tabular}{|l|c|c|c|c|}
\hline
\textbf{Model} & \textbf{Precision} & \textbf{Recall} & F1-score & \textbf{Accuracy (\%)} \\ \hline
RF & 0.85 & 0.85 & 0.85 & 88.8 \\ 
SVM & 0.82 & 0.86 & 0.84 & 84.3 \\ 
KNN & 0.64 & 0.52 & 0.68 & 64.0 \\ \hline
\end{tabular}
\caption{Comparison of classification models.}
\label{tab:comparison}
\end{table}

\begin{figure}[htbp]
    \centering
    \includegraphics[width=0.9\columnwidth]{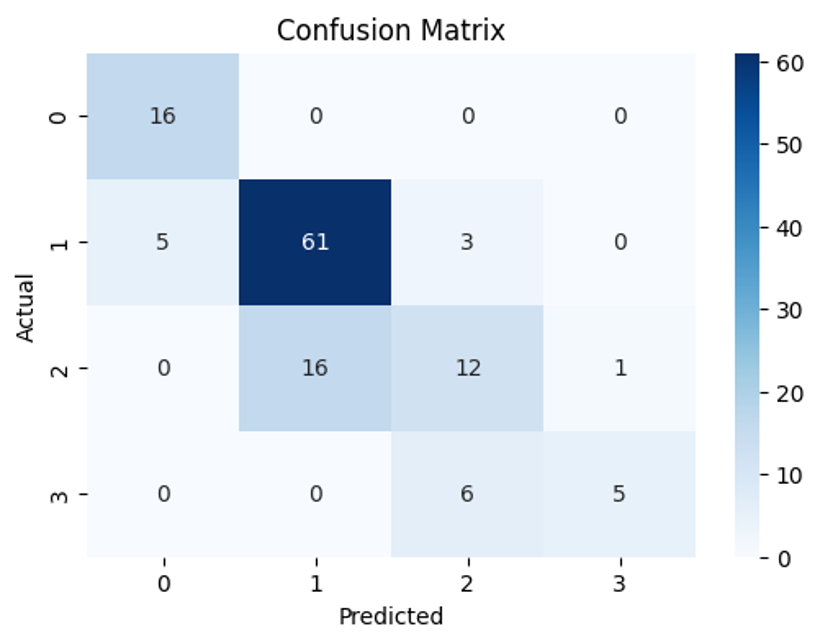}
    \caption{Confusion matrix of the Random Forest classifier.}
    \label{fig:conf}
\end{figure}

The confusion matrices in Figure \ref{fig:conf} revealed that Random Forest maintained a balanced prediction across all classes. These findings support Random Forest as the most robust choice for career prediction in this context.

\subsection{Comparison with Traditional Career Prediction Methods}
Traditional career prediction methods, which typically rely on resumes, academic qualifications, and interviews, often fail to capture real-time problem-solving abilities or technical skills. In contrast, our machine learning model evaluates quantifiable data such as user ratings, contest participation, and problem-solving trends, offering a more objective and data-driven approach.

While this approach has clear advantages - such as identifying hidden talent and reducing subjective bias, it also raises ethical considerations. Solely relying on competitive programming metrics may disadvantage individuals with limited access to such platforms or undervalue critical soft skills, such as teamwork and communication. To mitigate this, the model should ideally complement, rather than replace, holistic evaluation methods.

By leveraging these benefits while remaining mindful of its limitations, the machine learning-based approach presents a more effective and scalable method for identifying top programming talent. This comparison underscores the potential of data-driven models to enhance, rather than wholly redefine, career prediction and hiring processes.

\section{Case Studies on Real Codeforces Users}
This section demonstrates the practical application of the model by analyzing the career predictions for real Codeforces users. The case studies demonstrate how the model assesses competitive programming metrics to predict career status, offering insights into the growth trajectory and potential opportunities for each user.

\begin{figure}[htbp]
    \centering
    \includegraphics[width=0.5\textwidth]{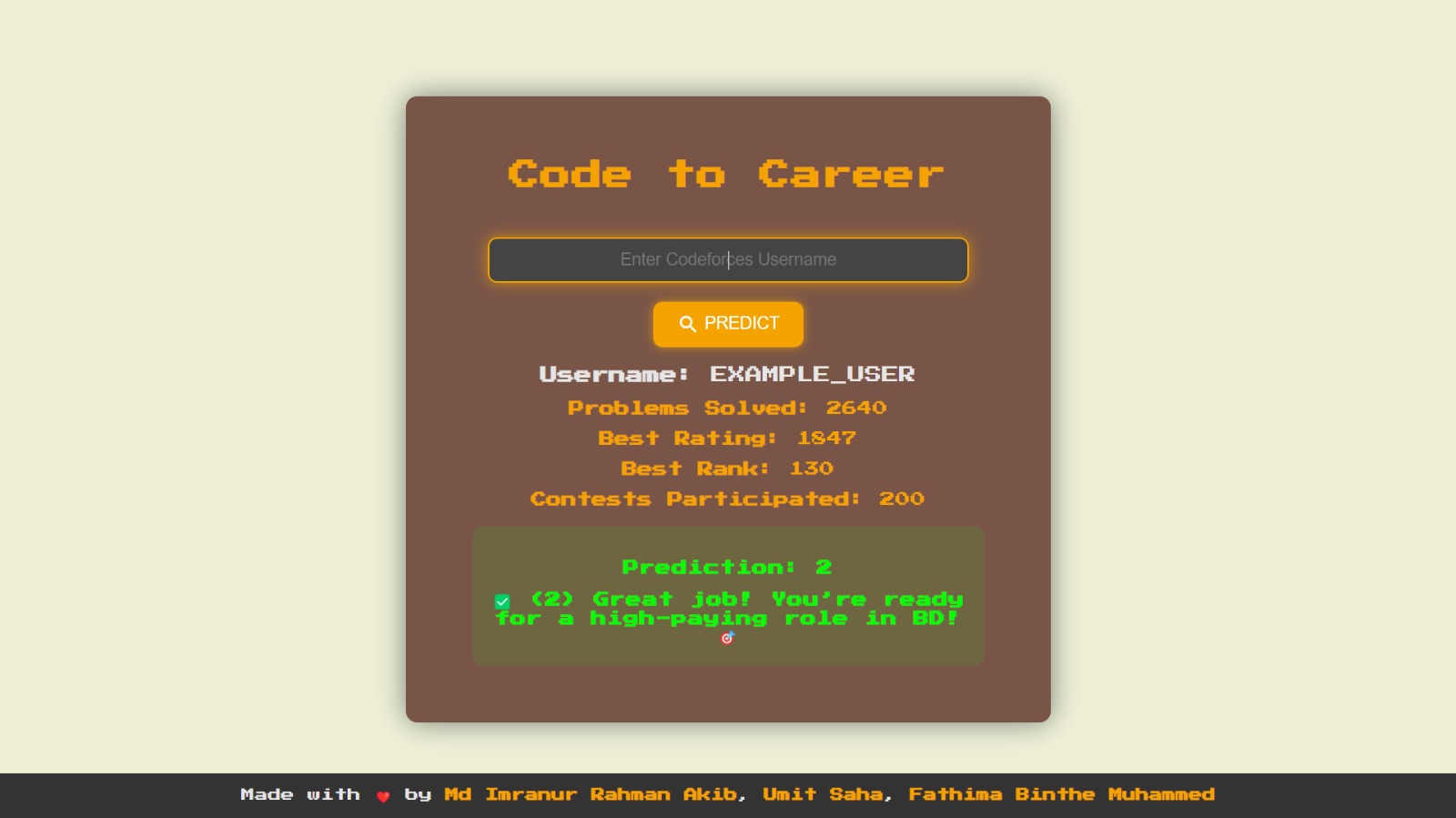} 
    \caption{Example of an user interface.}
    \label{fig:figure1}
\end{figure}

\subsection{User 1: "EXAMPLE\_USER\_1"}
The user "EXAMPLE\_USER\_1" has the highest rating of 1082, has solved 28 problems, and participated in 10 contests. The model predicted their career status as \textbf{0: Needs Improvement}. Despite the lower rating, the user's regular participation in contests reflects a commitment to skill development. The analysis suggests that, with continued effort and practice, they have the potential to improve their career prospects significantly.

\subsection{User 2: "EXAMPLE\_USER\_2"}
The user "EXAMPLE\_USER\_2" has the highest rating of 1847, has solved 2616 problems, and has participated in 200 contests. The model predicted their career status as \textbf{2: Ready for mid-level positions}. This user consistently demonstrates strong problem-solving skills and active participation in contests, indicating significant potential. The model identifies them as a strong candidate for competitive job roles, highlighting their persistence and engagement.

\subsection{User 3: "EXAMPLE\_USER\_3"}
The user "EXAMPLE\_USER\_3" has the highest rating of 2046, has solved 1149 problems, and participated in 176 contests. The model predicted their career status as \textbf{3: Ready for top tech companies}. With an impressive competitive programming profile, including a high rating and extensive problem-solving experience, this user is well-positioned to secure roles in top-tier tech companies. The model's prediction underscores their strong track record and potential for success in the tech industry.

These case studies illustrate how the model evaluates key performance indicators such as user ratings, problems solved, and contest participation to assess career potential. By analyzing real-world data, the model provides valuable insights into a programmer's growth trajectory and possible career opportunities, demonstrating its practical applicability in the tech industry.

\section{Conclusion}
This study confirms the potential of leveraging competitive programming data from Codeforces to predict career outcomes using machine learning. Among the models tested, the Random Forest classifier achieved the best performance by effectively utilizing features such as contest participation frequency, problem-solving diversity, and rating progression. To demonstrate practical applicability, a real-time prediction system was developed using a Flask-based web application and deployed via Render.com, highlighting how AI-driven tools can support career guidance and recruitment efforts.

Despite promising results, the study has limitations. These include dependence on partial Codeforces API data, a narrow focus on competitive programming metrics, and concerns around model overfitting and scalability. Future work will address these issues by incorporating data from other platforms like LeetCode and AtCoder, and integrating additional indicators such as GitHub activity or LinkedIn profiles to enrich career prediction. Furthermore, exploring advanced models like LSTMs or Transformers may enhance the system’s ability to learn temporal patterns. Enhancements such as personalized feedback, mobile accessibility, and migration to scalable cloud services like AWS or Google Cloud will also be pursued to improve usability and reach. With continued development, this system can evolve into a robust tool for aligning programmers' skills with career opportunities.

\bibliographystyle{IEEEtran}

\end{document}